\documentclass[aps,amsmath,preprint,a4paper,jcp]{revtex4}
\usepackage{graphicx}
\usepackage{amsmath}
\usepackage[T1]{fontenc}

\begin{document}


\title{First Passage Times for a Tracer Particle in Single File Diffusion and Fractional Brownian Motion}


\author{Lloyd P. Sanders}
\email[Corresponding author: ]{lloyd.sanders@thep.lu.se}
\affiliation{Department of Astronomy and Theoretical Physics, Lund University, SE-223 62 Lund, Sweden}

\author{Tobias Ambj\"ornsson}
\affiliation{Department of Astronomy and Theoretical Physics, Lund University, SE-223 62 Lund, Sweden}


\date{\today}

\begin{abstract}
We investigate the full functional form of the first passage time density (FPTD) of a tracer particle in a single-file diffusion (SFD) system whose population is: (i) homogeneous, \emph{i.e.} all particles having the same diffusion constant and (ii) heterogeneous, with diffusion constants drawn from a heavy-tailed power-law distribution. In parallel, the full FPTD for fractional Brownian motion [fBm - defined by the Hurst parameter, $H\in(0,1)$]  is studied, of interest here as fBm and SFD systems belong to the same universality class. 

Extensive stochastic (non-Markovian) SFD and fBm simulations are performed and compared to two analytical Markovian techniques: the Method of Images approximation (MIA) and the Willemski-Fixman approximation (WFA). We find that the MIA cannot approximate well any temporal scale of the SFD FPTD. Our exact inversion of the Willemski-Fixman integral equation captures the long-time power-law exponent, when $H \geq 1/3$, as predicted by Molchan [1999] for fBm. When $H<1/3$, which includes homogeneous SFD ($H=1/4$), and heterogeneous SFD ($H<1/4$), the WFA fails to agree with any temporal scale of the simulations and Molchan's long-time result. 

SFD systems are compared to their fBm counter parts; and in the homogeneous system both scaled FPTDs agree on all temporal scales including also, the result by Molchan, thus affirming that SFD and fBm dynamics belong to the same universality class. In the heterogeneous case SFD and fBm results for heterogeneity-averaged FPTDs agree in the asymptotic time limit. The non-averaged heterogeneous SFD systems display a lack of self-averaging. An exponential with a power-law argument, multiplied by a power-law pre-factor is shown to describe well the FPTD for all times for homogeneous SFD and sub-diffusive fBm systems.
\end{abstract}


\maketitle


\section{Introduction}
Within the physical phenomena that exhibit stochasticity, the idea of the first passage time density (FPTD) is of great importance \cite{Redner2001}. Recently the literature has seen analysis of FPTDs in a wide variety of research areas, including stochastic systems: Diffusion with stochastic resetting \cite{Evans2011}, diffusion in complex scale-invariant media \cite{Condamin-2007}; cosmology (the calculation of the mass distribution of dark matter halos) \cite{maggiore-2010}; evolutionary ecology (foraging tactics of various animal species) \cite{Benichou-2005}; and covered extensively in primarily one-dimensional (1D) single particle scenarios by Redner \cite{Redner2001}. Most previous studies focus on FPTD for Markovian (memory-less) dynamics.

The role of crowding in physical systems is of key relevance in providing non-Markovian effects in many scenarios, especially within the cell (e.g. \cite{Ellis:2001, Banks2005, Golding-2006}). An important special case in this area of stochastic processes is single-file diffusion (SFD), defined as many particles diffusing in a 1D system, where the order of the particles within the system remains constant, \emph{i.e.} the particles are subject to hard-core repulsion. Experimentally SFD is shown to exist in many types of systems, such as the diffusion of colloids in channels \cite{Wei-2000,Lutz-2004}, and of relevance in the motion of fluorescently tagged proteins on DNA \cite{LippincottSchwartz-2009}.

Initially, because of the problem's mathematical tractability; over the past five decades, beginning with the hallmark theoretical investigations of Harris \cite{harris-65}, followed by Lebowitz and Percus \cite{Lebowitz1967}, and, for instance, Fisher \cite{Fisher1984}, a large amount of (theoretical) work has been conducted into SFD systems, notably in the latter decade \cite{Rodenbeck-98, lizana-2009, lizana-2009-80, lomholt-2010, lizana-2008-100, taloni-lomholt-2008}, generalizing the original results to particles interacting with general potentials and in external force fields. Of interest for  the present study is the recent increasing amount of research into complex population types or heterogeneous SFD systems \cite{ambjornsson-2008-129, lomholt-2010, Jara2008, Jara2009, Flomenbom2010}.

One prominent example where SFD and the FPTD are amalgamated is within the context of protein-DNA interaction. It was shown that the time taken for a protein to find a specific start location is up to two orders of magnitude greater than predicted by Smoluchowski three-dimensional (3D) diffusion rate alone \cite{Hippel-1989}. This led to the concept of ``facilitated'' target location - the combination of 3D diffusion through the cytoplasm and 1D diffusion along the DNA; this coupling is shown to exist experimentally \cite{Sokolov2005, Wang2006, Elf2007}, and corroborated theoretically \cite{Li-2009}. 

Due to the importance of SFD FPTD in biological (and other) systems, there is a surprising lack of research into the explicit nature of the FPTD for a tracer particle of a SFD system, but a few exceptions exist, for example Li \emph{et al.} \cite{Li-2009} used the Method of Images \cite{Jackson} to help explain the FPT of proteins on ``road-blocked'' DNA strands in conjunction with 3D diffusion. Taloni \emph{et. al.} \cite{taloni-10}, performed homogeneous SFD simulations, with particular emphasis on the long-time asymptotics. 

In the field of stochastic processes there exists a phenomenon known as anomalous diffusion \cite{Metzler2000, Metzler2004, Weigel2011, Jeon2011B} whose ensemble-averaged mean-squared displacement is represented by non-linear time dependence: $\langle \mathbf{x}^2(t)\rangle \sim t ^{2H}$, where $2H\neq1$.  A prominent example of anomalous diffusion is the process of fractional Brownian motion (fBm) \cite{Kolmogorov1940, Mandelbrot1968}. fBm is defined by a single parameter, the Hurst exponent $H \in (0,1)$; and is a generalized, zero mean, stationary, Gaussian process with increments obtained from fractional Gaussian noise (fGn), producing a position auto-correlation function \cite{Qian2003}: $\langle B_H(t)B_H(s)\rangle=C\left( t^{2H}+s^{2H}-\mid t-s\mid^{2H} \right)$, where $C$ is a constant. In terms of displacement, a fBm time series is represented by the cumulative sum of the fGn, $\eta(t)$, whose auto-correlation function is $\langle\eta(t)\eta(0)\rangle = H\left(2H-1\right)|t|^{2H-2}+2\left(1+2H\right)\delta(t)^{2-2H}$, see \cite{Qian2003}; where it is interesting to note that this function for $t>0$ is equivalent to the memory kernel of the fractional Langevin Equation (fLE) \cite{KouXie-2004, lizana-2009, Fluc_diss_note}. When $H<1/2$, the pre-factor in the fGn auto-correlation function becomes negative for $t>0$, illustrating an anti-correlated process \cite{fGn}. fBm is of relevance in fields such as: Hydrology \cite{Hurst1951} (the investigation from which the Hurst exponent derives its name), diffusion of biomolecules \cite{Weber2010}, quantitative finance \cite{Cont2001} and even considered in genetic algorithms \cite{Jara2011}. It should be noted that there are other processes which can give rise to anomalous-type diffusion, two of which are covered extensively within the literature: Continuous-time random walks \cite{Klafter1987}; and diffusion on fractals \cite{Nakayama1994}. For a comprehensive overview of anomalous diffusion see the reports by Metzler \cite{Metzler2000}, and later, Metzler and Klafter \cite{Metzler2004}.

The full functional form for the FPTD of a fBm process has remained elusive, but recent work into various FPT scenarios has become common of late. Such as, escape problems with fGn \cite{KramerEscape2010}, and FPT of fBm in 2D wedge domains \cite{Jeon2011}. In 1D, it was proposed by Ding and Yang \cite{Ding1995}, shortly after, a simple, succinct, physical argument was proposed by Krug and co-authors \cite{Krug1997}, and later rigorously proven by Molchan \cite{molchan-99}, that the long-time form for the FPTD is 
\begin{equation}\label{eq:Molchan}
 f(t)\simeq t^{H-2}.
\end{equation}
This result is a corner stone of our investigation.

It is well-known in the mathematics literature that tracer particle dynamics in a hard-core lattice gas (symmetric exclusion process) is equivalent to fBm, see for instance \cite{Jara2008} (see Refs. \cite{lizana-2009, taloni-10} for the case of continuum motion). Thus these processes/FPTDs are intimately linked via the theoretical long-time benchmark that is Molchan's result, Eq. (\ref{eq:Molchan}). It was demonstrated in Ref. \cite{lizana-2009} that a homogeneous SFD system can be modeled with a Hurst parameter equaling $H=1/4$, in the asymptotic time limit. We here go beyond previous studies, using Eq. (\ref{eq:Molchan}) as guidance, by examining the \emph{full} functional form of the FPTD, for SFD (homogeneous and heterogeneous) and fBm systems for all times, by extensive simulations and analytic approximations; so as to attempt to elucidate the entire FPTD relationship between these systems and the universality class to which they both belong.

The layout of this article is as follows: In $\S$ \ref{sec: Approx}, the SFD tracer particle/fBm probability density function (PDF) is presented, and using this result, two Markovian techniques - Method of Images (MIA) and Willemski-Fixman (WFA) - are used to approximate the FPTD of the tracer particle within homogeneous and heterogeneous SFD systems, including also the FPTD of a fBm particle. The results are analyzed and compared to our simulations in the penultimate section. Whereas the MIA was used before, our explicit analytical inverse of the so called Willemski-Fixman relation is to our knowledge, new. Within this section, a simple functional form is conjectured to model the FPTD for all times and for all $H$. $\S$ \ref{sec: Sims} describes how the SFD (homogeneous and heterogeneous) and fBm systems are simulated and what sets of parameters are used. Previous theoretical results for tracer particle dynamics are briefly reviewed. In $\S$ \ref{sec: Results} the results concerning all aspects of the investigation are discussed in detail; this leads to $\S$ \ref{sec: conclu} that concludes the results herein, with a brief discussion on proposed further future investigations based upon this work. The more technical details regarding the derivations of the FPTD for both approximations are left for the Appendices \ref{sec:WFA} and \ref{sec:MIA}. Data fitting procedures are situated in Appendix \ref{sec:DataRed}.

\section{Approximating the FPTD\label{sec: Approx}}
To investigate the FPTD of tracer particle dynamics in SFD and fBm systems, which are inherently non-Markovian, we begin by analyzing the applicability of two well known Markovian methods; both of which require the following results: The tracer particle PDF for SFD systems and the fBm particle PDF is Gaussian \cite{harris-65} (for infinite systems):
\begin{equation}\label{eq:GeneralPDF}
P(x,t|x_0)=\frac{1}{\sqrt{2\pi S(t)}}\exp\left(-\frac{ (x-x_0)^2}{2S(t)}\right), 
\end{equation}
where $x_0$ is the position of the tracer particle at the initial time, and $x$ the position. The mean-squared displacement (MSD) is denoted $S(t)$ and is, in the long-time for tracer particle dynamics in SFD systems and of a fBm particle \cite{KramerEscape2010}, 
\begin{equation}\label{eq:MSDfBm}
S(t)\equiv \langle\left[x(t)-x_0\right]^{2}\rangle = 2Ct^{2H}.
\end{equation}
The angled brackets denote the ensemble average \cite{Time_av_MSD}, and $C$ is the effective diffusion constant. The explicit expression for $C$ for SFD systems is discussed in $\S$ \ref{sec: Sims}. For fBm, $C$ is the amplitude of the process (here as position) auto-correlation function, although often set to $C=1/2$ in the literature \cite{Qian2003}. With these two expressions we can now develop two approximations for the FPTD.

\subsection*{The Willemski-Fixman Approximation}
In a Markovian system, one is able to acquire the PDF and in turn, the FPTD through the well-known Willemski-Fixman (also known as Renewal Theory) method \cite{VanKampen}. Simply put, a particle must have a first passage through point $x_c$ at some time $t'$ to reach point $x$ at time $t\geq t'$, \emph{i.e.}, functionally as
\begin{equation}\label{eq:renewal}
 P(x,t|x_0) = \int_0^{t}f(x_c,t'|x_0)P(x,t-t'|x_c)dt',
\end{equation}
which constitutes a convenient integral equation for the FPTD, $f(x_c,t|x_0)$, if one sets $x=x_c$ in Eq. (\ref{eq:renewal}). However, due to the power-law argument with respect to time in the exponent in Eq. (\ref{eq:renewal}) for the PDF, Eq. (\ref{eq:GeneralPDF}), we cannot use the usual Laplace-transform technique for inverting Eq. (\ref{eq:renewal}) \cite{VanKampen}. However, through the use of Mellin transforms \cite{IntTransBook}, we are able to follow the usual line of attack in this type of problem: transform the convolution into a product in ``frequency'' (Mellin) space $p$, rearrange to find the Mellin-transformed FPTD $\hat{f}(x_c,p|x_0)$, then transform back to $t-$space and re-acquire $f(x_c,t|x_0)$. The explicit derivations are left to Appendix \ref{sec:WFA}, and the final results are given here:

Using a series expansion approach \cite{Hughes} of our exact expression for $\hat{f}(x_c,p|x_0)$, we find an exact expression for the FPTD within the WFA for the general case ($0<H<1$; with $H\neq H^{*}_{nm}$, see below) through the inversion of the Mellin transform. Explicitly, 
\begin{align}\label{eq: series-exp}
 f_{\textrm{WF}}(x_c,t|x_0)&=&\frac{\sigma^{-1/(2H)}}{\Gamma(1-H)}\bigg{[}\left[K(t)\right]^{2-H}\frac{1}{2H}\sum^{\infty}_{n=0}\left[K(t)\right]^{n}\frac{(-1)^{n}}{n!}\frac{\Gamma\left(\frac{H-1-n}{2H}\right)}{\Gamma\left(H-1-n\right)} \nonumber \\
&+& K(t)\sum^{\infty}_{m=1}\left[K(t)\right]^{2Hm}\frac{(-1)^{m}}{m!}\frac{\Gamma\left(1-H-2Hm\right)}{\Gamma\left(-2Hm\right)}\bigg{]}
\end{align}
where $\sigma=(\Delta x)^2/4C$, with $\Delta x = x_c-x_0$, $\Gamma(z)$ is the Gamma function \cite{ABST}, and $K(t)$ is defined as 
\begin{equation}\label{eq:WFK}
 K(t)=\frac{\sigma^{1/2H}}{t}.
\end{equation}
Eq. (\ref{eq: series-exp}) is very computationally efficient for calculating FPTDs, but is only valid when $H$ is \emph{not} a rational number of the form $H^{*}_{nm}=(n+1)/(2m+1)$ where $n$ and $m$ are positive integers (see Appendix \ref{sec:WFA} for further details). Eq. (\ref{eq: series-exp}) generalizes the long-time expression in \cite{Bologna2010} to include also shorter time-scales. Eq. (\ref{eq: series-exp}) is equivalent to the usual Brownian motion result when $H=1/2$; again see Appendix \ref{sec:WFA}. Alternatively, the inversion back from Mellin-space can also be completed using Weyl fractional derivatives \cite{Metzler2000, IntTransBook}, which yields an equivalent expression 
\begin{equation}\label{eq:WFf}
 f_{\textrm{WF}}(x_c,t|x_0)=\frac{2\sin(\pi H)}{\pi}\sigma^{-1/(2H)}[K(t)]^{(2-H)}I[K(t)],
\end{equation}
\begin{equation}\label{eq:WFI}
 I(z)=\int_0^\infty y^H[2H(z+y)^{2H}+1-2H](z+y)^{2H-2}\exp{\left[-(z+y)^{2H}\right]}dy,
\end{equation}
valid except for $K(t)\rightarrow0$ for $H\leq1/3$. 

To check the validity of the WFA, we look toward the long-time limit, $t\gg\sigma^{1/(2H)}$, for comparison to the theoretical result given by Eq. (\ref{eq:Molchan}). In this limit where $K(t)\rightarrow 0$, the series expansion result gives:
For $H>1/3$
\begin{equation}\label{eq:SE H>1/3}
 f_{\textrm{WF}}(x_c,t|x_0)\sim\frac{\sin(\pi H)}{\pi}\sigma^{-1/(2H)}\Gamma\left(\frac{3H-1}{2H}\right)[K(t)]^{(2-H)}\propto t^{H-2}.
\end{equation}
For $H<1/3$
\begin{equation}\label{eq:SE H<1/3}
 f_{\textrm{WF}}(x_c,t|x_0)\sim\frac{2H\Gamma\left(1-3H\right)}{\Gamma\left(1-H\right)\Gamma\left(1-2H\right)}\sigma^{-1/(2H)}[K(t)]^{(1+2H)}\propto t^{-1-2H}.
\end{equation}
When $H=1/3$, the long-time limit becomes 
\begin{equation}\label{eq:SE H=1/3}
 f_{\textrm{WF}}(x_c,t|x_0)\sim\frac{\sqrt{3}}{2\pi}\left(1+\frac{5}{3}\gamma_E\right)\sigma^{-3/2}[K(t)]^{5/3}\propto t^{-5/3}.
\end{equation}
as shown in Appendix \ref{sec:WFA}. Euler's constant \cite{ABST} is given as $\gamma_E$. Note that our long-time results coincide with Bologna et al. \cite{Bologna2010}; but therein, the authors solve only up to a pre-factor, whereas our result provides a general explicit solution including the pre-factor. Combining Eqs. (\ref{eq:SE H>1/3}), (\ref{eq:SE H<1/3}), and (\ref{eq:SE H=1/3}) gives the FPTD within the WFA for all $H$ in the asymptotic limit. Interestingly,  when $H\geq 1/3$, as in Eqs. (\ref{eq:SE H>1/3}) and (\ref{eq:SE H=1/3}) the WFA approach is in agreement with Molchan's result. However, for the case that $H<1/3$, Eq. (\ref{eq:SE H<1/3}) shows a different exponent to Molchan's $(H-2)$. Thus our work shows that $H<1/3$ is, in a sense, ``too far'' from the Brownian case ($H=1/2$) for renewal theory to apply. Further analysis of these results are left to $\S$ \ref{sec: Results}.

\subsection*{The Method of Images Approximation}
The Method of Images is a technique used in calculating properties of electrostatics \cite{Jackson}, but has also been applied to Markovian stochastic systems \cite{Redner2001}. This technique calculates the resultant PDF from having a PDF in the absence of a boundary, centered at $x_0$, and an identical PDF centered at $2x_c-x_0$, which is inverted. The result ``creates'' the existence of a boundary at $x_c$; from this, the FPTD can be calculated. For the case of non-Markovian systems, such as fBm or the SFD system, this method is only an approximation, due to the fact that the state of the system is dependent upon all previous trajectories; namely the system is \emph{not} memory-less. Never-the-less, this technique has been used in the literature to describe the FPTD in similar situations \cite{Li-2009}. The resultant FPTD (see Appendix \ref{sec:MIA} for derivation) is
\begin{equation}\label{eq:MIApprox}
 f_{\textrm{MI}}(x_c,t|x_0)=\frac{H\Delta x}{\sqrt{\pi C}t^{H+1}}\exp\left(-\frac{\Delta x^2}{4Ct^{2H}}\right).
\end{equation}
The FPTD for a Brownian particle is produced when $H=\frac{1}{2}$ and $C=D$, where $D$ is the diffusion constant, as it should. However, for any system such that $H\neq1/2$, the long-time behavior predicted by the MIA, $f_{\textrm{MI}}(t)\sim t^{-(H+1)}$, where this result and  Eq. (\ref{eq:Molchan}) are in clear conflict. A more in-depth analysis of the MIA with respect to simulations is left to $\S$ \ref{sec: Results}.

\subsection*{Conjecture for the FPTD\label{sec:Guess}}
As we will see in $\S$ \ref{sec: Results} the Markovian approximations fail to produce the correct full functional form FPTD for the two physical SFD systems present here, and more explicitly, analytically, only the WFA appears to capture the theoretical long-time result for $H\geq 1/3$. Therefore we here provide a simple conjecture made in light of the results in previous work and more simplistic systems to account for the full form. 

Our proposed function for the FPTD has two main features: (i) a function which becomes exact for the Brownian case and (ii) approaches the correct long-time behavior for both SFD and fBm systems, Eq. (\ref{eq:Molchan}), for arbitrary $H$ (which, obviously, includes those values which represent SFD homogeneous and heterogeneous systems). Namely 
\begin{equation}\label{eq:fptGuess}
 f_{\textrm{c}}(x_c,t|x_0)=\Omega t^{H-2} \exp\left(-\gamma\left[\frac{\Delta x^2}{2Ct^{2H}}\right]^{\beta}\right),
\end{equation}
with dimensionless fitting parameters $\gamma$ and $\beta$. The normalization constant, $\Omega$, is 
\[
\Omega= \frac{2H\beta}{\Gamma\left(\frac{1-H}{2H\beta}\right)}\left[\left(\frac{\Delta x^2}{2C}\right)^{\beta}\gamma\right]^{\frac{1-H}{2H\beta}}.
\]
Consideration of this conjecture in conjunction with simulations is left to $\S$ \ref{sec: Results}.
\section{Simulations\label{sec: Sims}}

\begin{figure}[!ht]
\centering
\includegraphics[scale=0.6]{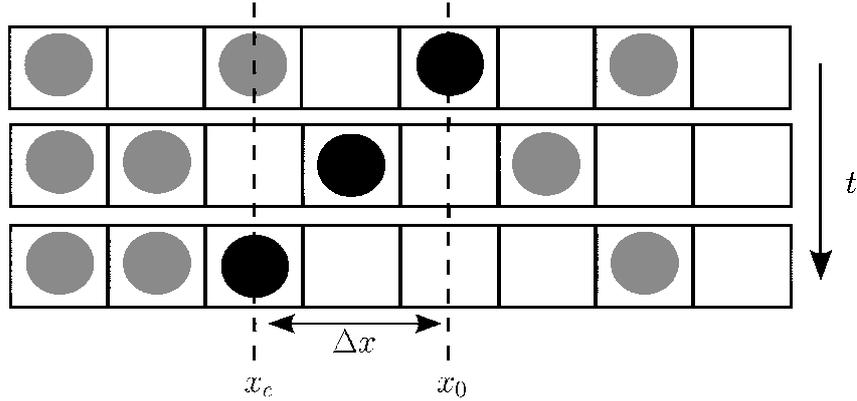} 
\caption{Schematic representation of the lattice simulations of a SFD system. All particles (including the tracer - here in black) move under Brownian motion and are hard-core (the particles cannot occupy the same lattice site), meaning that they cannot pass each other, keeping their order for all time. Hence the tracer is in the center of all other particles for all times. The top panel shows the start of the simulation in thermal equilibrium, whereas the bottom panel shows that after some time, the tracer has achieved a first passage event.\label{fig:Lattice}}
\end{figure}

{\bf Simulating SFD:} To computationally simulate the FPTD in a single-file system, the Gillespie-type algorithm for hard-core lattice dynamics presented in Ref. \cite{ambjornsson-2008-129} is implemented. Stochastic time series are generated for both homogeneous and heterogeneous population types through the following steps: (1) Place the particles in their initial, thermally equilibrated, positions (the tracer particle is positioned in the middle of the system with equally many particles randomly distributed to the left and the right). (2) Move a random particle according to the algorithm in \cite{ambjornsson-2008-129} and update the time $t$. (3) repeat step (2) until either $t\geq t_{\textrm{stop}}$ (some designated stop time) or the tracer particle position becomes $\geq x_c$, and record the corresponding first-passage time; see Fig. \ref{fig:Lattice} for a diagrammatic explanation.

Both homogeneous and heterogeneous particle populations were simulated using the parameters provided in Table \ref{table:SFD_param}. The simulation box size, $L$, was chosen such that the results accurately display the dynamics for a semi-infinite system given the finite nature of a computer simulation \cite{FiniteSystem}. 
\begin{table}[ht]
\caption{SFD Simulation Parameters. Not applicable is abbreviated to N.A.}
 \centering
\begin{tabular}{|c|c|c|}
	\hline
Parameter & Homogeneous & Heterogeneous\\
 & systems & systems\\
	\hline
System Size, $L$ (lattice site width $a$) & $10^4$ $a$& $2\times10^4$ $a$\\
Particle Number Density, $\rho$& 0.25& 0.25\\
Diffusion Constant, $D$& $1$ $a^2\textrm{s}$& N.A.\\
Average Diffusion Constant, $\bar{D}$& N.A.& $1$ $a^2\textrm{s}$\\
Stop time, $t\geq t_{\textrm{stop}}$ & $2.5\times10^7$ $a^2/D$& $10^8$ $a^2/\bar{D}$\\
Friction constant PDF exponent, $\alpha$& N.A.& $0.5$\\
Initial tracer particle position, $x_0$ & $0$& $0$\\
Simulation Ensemble size, $N$ & $2.5\times10^3$& $2.5\times10^3$\\
	\hline
\end{tabular}
\label{table:SFD_param}
\end{table}
The lattice simulations conducted have focused on one particle density ($\rho=0.25$), one particle size, $a$ (which is equivalent to the lattice site length), and four different absorption point differences, $\Delta x$ (shown in Table \ref{table:GuessFitHom}). 

For heterogeneous SFD systems the particles' friction constants ($\xi \propto D^{-1}$ in units of thermal energy; $D$ is the diffusion constant) are drawn from a so-called heavy-tail (HT) power-law distribution (the first moment is not finite) \cite{lomholt-2010, LT_fric_const}. The HT frictional constant distribution is chosen as $\varrho(\xi)=\Xi\xi^{-1-\alpha}$, for large $\xi$, where $0<\alpha<1$ and $\Xi$ is a normalization constant. For such heterogeneous SFD systems the dynamics are found to be inherently slow, and therefore large systems and simulation stop times are required (compared to the same system with a homogeneous population), this is accounted for in our study, as shown in Table \ref{table:SFD_param}.

{\bf Simulating fBm:} We here use the so-called Spectral simulation method; as it uses the spectral density of the fGn increments to calculate the time series. This method was chosen for its speed (use of fast Fourier transforms) and simplicity of implementation, which approximates well fBm, for long series \cite{Dieker2003}. For sub-diffusive systems with slow dynamics, a relatively large (with respect to SFD systems herein) effective diffusion constant, $C = 5$, is used to efficiently calculate and display all relevant time frames in a reasonable computation time. See Table \ref{table:fBm_params} for all fBm simulation parameters \cite{Approx_sim}, and results.

{\bf Collapsing Data:} When comparing sets of data from the same system (characterized by $H$) with different absorption points and/or generalized diffusion constants, it is convenient to collapse the data to remove these dependencies on $x_c$ and $C$ and thereby be able to extract a universal functional form for the FPTD. This is done through knowledge of length versus time scaling in anomalous diffusion, provided by Eq. (\ref{eq:MSDfBm}). Through observation of this equation it is convenient to introduce  a scaling factor $\varpi = \left(\Delta x\right)^{-1/H}C^{1/2H}$ with dimension inverse time, and to scale FPTDs with this factor, as it is done in Figs. \ref{fig:HomR25Collapse} to \ref{fig:fBmH75Collapse}. 

For homogeneous SFD systems it has been shown \cite{lizana-2009-80} that the MSD for a tracer particle is described by, see Eq. (\ref{eq:MSDfBm}),
\begin{equation}\label{eq:MSDboxHom}
 C \sim\frac{(1-\rho a)}{\rho}\sqrt{\frac{D}{\pi}},\quad H=\frac{1}{4},
\end{equation}
where $D$ is the single particle diffusion constant, and $a$ is the size of the particle. For heterogeneous SFD systems Lomholt \emph{et al.} \cite{lomholt-2010} find the approximate long-time MSD-pre-factor and Hurst exponent to be 
\begin{equation}\label{eq:MSDboxHet}
 C \sim\frac{k_{B}T}{2\sqrt{\kappa \chi}}\frac{1}{\Gamma(1+\delta)}, \quad H=\frac{\alpha}{2(1+\alpha)},
\end{equation}
where $\kappa=\rho^2 k_BT(1-\rho a)^{-2}$, $\chi=(4\kappa)^{1-4H}(\Xi\pi/\sin[2H\pi/(1-2H)])^{2(1-2H)}$, and $\alpha$ is the exponent in the HT friction PDF. Note that the Hurst exponent satisfies $H<1/4$ in these heterogeneous systems, as $0<\alpha<1$; we use $\alpha=1/2$ during this investigation (see Table \ref{table:SFD_param}). From Eq. (\ref{eq:MSDboxHet}), $\varpi$ may be calculated for heterogeneous SFD systems, likewise for homogeneous systems using Eq. (\ref{eq:MSDboxHom}).

\section{Results\label{sec: Results}}
\begin{figure}[!ht]
\centering
\includegraphics[scale=1.2]{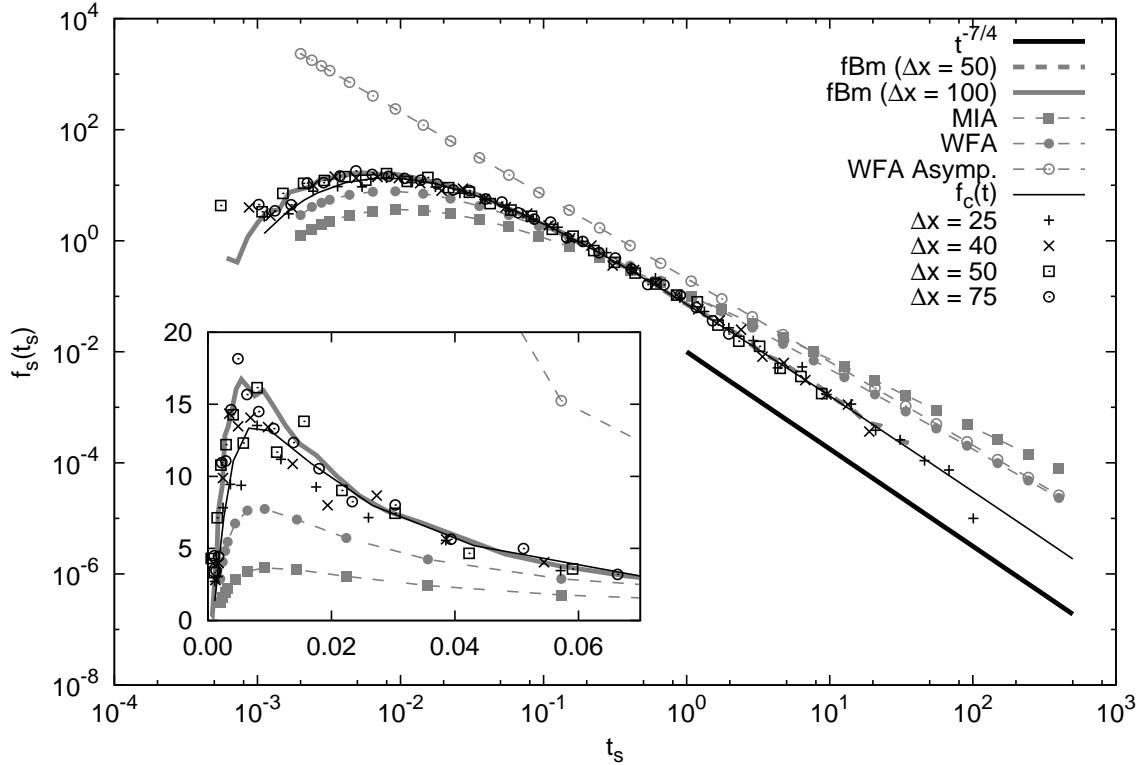} 
\caption{Collapsed (log-log) plot of FPTD for a \emph{homogeneous} SFD system with different absorption points and fBm FPTD (for simulation parameters see Tables \ref{table:SFD_param} and \ref{table:fBm_params}, respectively). Immediately, the collapsed plot shows that homogeneous SFD and fBm have the same FPTD dynamics over all time frames in the correct scaling, $\S$ \ref{sec: Sims}. The MIA, Eq. (\ref{eq:MIApprox}), and the WFA, Eq. (\ref{eq: series-exp}), are shown for comparison, both of which show poor agreement. The averaged proposed functional form, $f_{\textrm{c}}(t)$, Eq. (\ref{eq:fptGuess}), is constructed by collapsing all simulated data and fitting (see Appendix \ref{sec:DataRed}), keeping the power-law exponent in the pre-factor fixed to $H-2$; with $H=1/4$, and setting $C$ according to Eq (\ref{eq:MSDboxHom}). The parameters  (Table \ref{table:GuessFitHom}) were then averaged and a single mean curve plotted. This conjecture shows excellent agreement with both anomalous diffusive systems on all time scales \cite{fBm_hom_conjec_fit}. The Molchan long-time prediction \cite{molchan-99} is given to guide the eye. The fBm FPTD consists of two different absorption points ($\Delta x=50$, $\Delta x=100$, $6\times 10^4 $ simulations) \cite{fBm_overlap_hom} and displays good agreement with SFD results. INSET: The short-time regime (linear axes) agreement between homogeneous SFD, fBm, and $f_{\textrm{c}}(t)$. Remaining simulation details are presented in Tables \ref{table:SFD_param}, \ref{table:fBm_params}, and \ref{table:GuessFitHom}. The subscript $s$ on each axis variable denotes ``scaled'', namely $t_s=\varpi t$ and $f_s=\varpi^{-1}f(t)$, where $\varpi=C^{1/2H}\left(\Delta x\right)^{-1/H}$.\label{fig:HomR25Collapse}}
\end{figure}

\begin{figure}[!ht]
\centering
\includegraphics[scale=1.2]{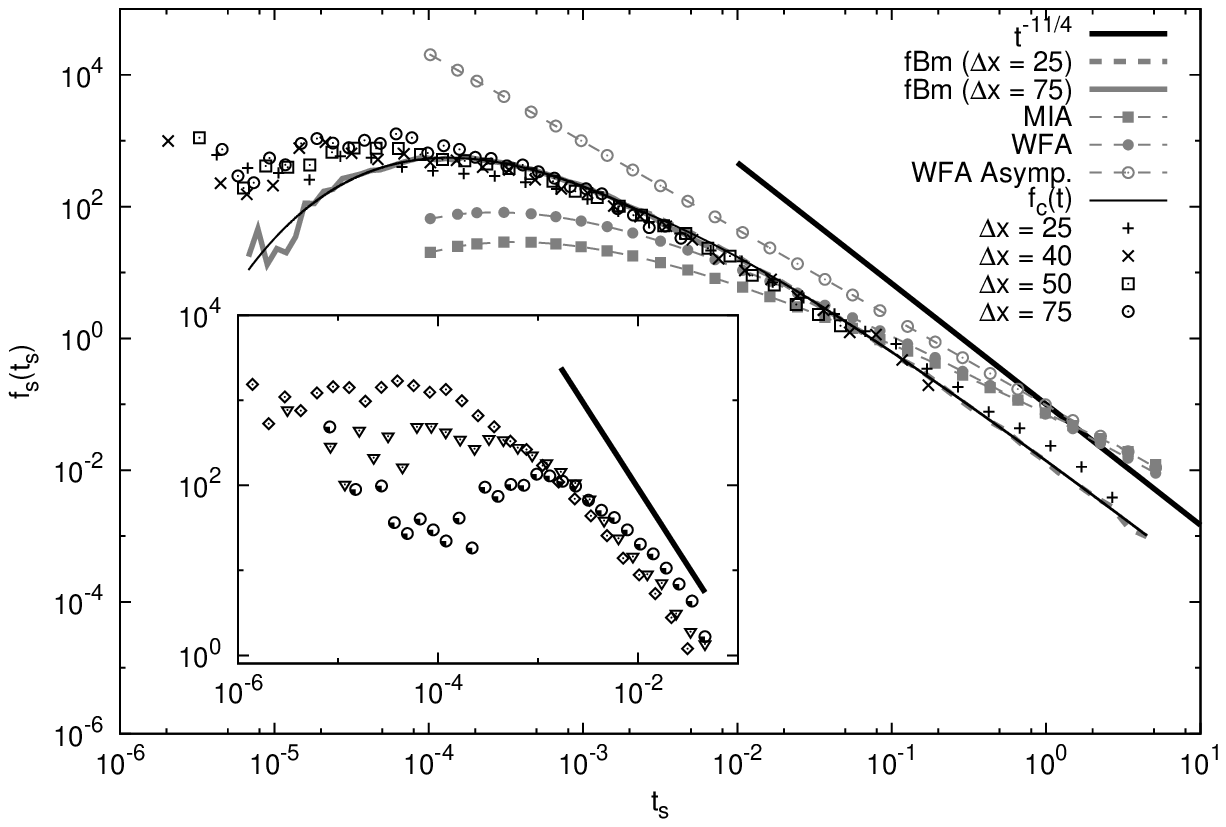} 
\caption{Collapsed plot of FPTD for a \emph{heterogeneous} SFD system (Table \ref{table:SFD_param}) with different absorption points. Also within, fBm FPTD is plotted for $H=1/6$, see Table \ref{table:fBm_params}. The long-time dynamics for both systems agree very well with each other and also with Molchan's equation, Eq. (\ref{eq:Molchan}). In the very short-time the systems part, most likely because of the complex nature of heterogeneity-averaged SFD systems (see inset). The MIA, Eq. (\ref{eq:MIApprox}), and the WFA, Eq. (\ref{eq: series-exp}), are shown for comparison. Both approximations show ill agreement with the simulated data for both systems and the theory. $f_{\textrm{c}}(t)$, Eq. (\ref{eq:fptGuess}), was fitted to the fBm data (displaying excellent agreement), Table \ref{table:fBm_params}, as opposed to the SFD data, due to its poor fit (discussion in inset caption here, and $\S$ \ref{sec: Results}) \cite{fBm_overlap_het}. The data is scaled (signified by subscript $s$) using Eq. (\ref{eq:MSDboxHet}), with $H=1/6$ (since $\alpha=1/2$) - see Fig. \ref{fig:HomR25Collapse} caption for further explanation of the scaling. All remaining simulation details are presented in Tables \ref{table:SFD_param} and \ref{table:fBm_params}. INSET: Collapsed plot of the FPTD for 3 different sets of heterogeneous friction constants, kept constant for each simulation. For this non-averaged case we use $x_c = 50$ for all simulations, with all other system parameters displayed in Table \ref{table:SFD_param}. The inset illustrates the fact that no self-averaging takes place in this heterogeneity-averaged system (see $\S$ \ref{sec: Results} for further discussion).\label{fig:HetR25Collapse}}
\end{figure}

{\bf SFD simulations:} Each different data set (defined by its unique absorption point, $x_c$) is shown in a collapsed plot in Fig. \ref{fig:HomR25Collapse} for homogeneous particles and Fig. \ref{fig:HetR25Collapse} for heterogeneous systems. There are concurrent motifs in both FPTD plots; namely the short-time concave characteristic maximum moving into a long-term power-law type structure. 
For the heterogeneous case, Fig. \ref{fig:HetR25Collapse}, there is more variance in the short-term time regime compared to the same time regime for the homogeneous, Fig. \ref{fig:HomR25Collapse}. This is because the system is given a new random set of diffusion constants for every simulation (the so-called heterogeneity-averaged case \cite{lomholt-2010}), therefore giving a super-position of non-averaged FPTDs (see discussion at the end of this section) as demonstrated in Fig. \ref{fig:HetR25Collapse} inset.

{\bf Long-time SFD asymptotics:} An important check of simulating homogeneous SFD data is that it should agree with Eq. (\ref{eq:Molchan}), for long times. Upon evaluation of Fig. \ref{fig:HomR25Collapse}, where Molchan's asymptotic result is placed to guide the eye, one sees that the data agrees well with theory, in accordance with previous literature findings \cite{taloni-10} for such a system, further confirming that SFD systems and fBm belong both to the same universality class in the asymptotic limit.

A novel result then was to verify if Molchan's work agreed with the heterogeneous SFD system proposed and investigated by \cite{lomholt-2010}, assuming $H$ is given by Eq. (\ref{eq:MSDboxHet}). In Fig. \ref{fig:HetR25Collapse}, in both the main and inset plots, the theoretical asymptotic result (which is again placed to guide the eye) agrees well with the long-time FPTD for heterogeneity-averaged simulations. 

{\bf SFD versus fBm:} In both collapsed plots (Figs. \ref{fig:HomR25Collapse} \cite{fBm_overlap_hom} and \ref{fig:HetR25Collapse}  \cite{fBm_overlap_het}) the fBm simulations agree well with the SFD data, including  the scaled power-law pre-factor. 

Comparison of the short-time regimes in Fig. \ref{fig:HomR25Collapse}, and in particular the inset, illustrates that the fBm and homogeneous system have the same FPTD on all temporal scales.

Within the short-time heavy-tailed heterogeneous SFD regime, although possessing the same structure, the fBm simulations do not coincide with the heterogeneous data (see discussion at the end of the section). 

{\bf SFD data versus approximations:} In the homogeneous system, as demonstrated in $\S$ \ref{sec: Approx}, the long-time MIA FPTD, Eq. (\ref{eq:MIApprox}), does not agree with Eq. (\ref{eq:Molchan}). With inspection of Fig. \ref{fig:HomR25Collapse} we note that although the concave structure of $f_{\textrm{MI}}(t)$ is present and similar to the data, these two FPTDs do not coincide, making for an obvious case that $f_{\textrm{MI}}(t)$ does not agree with the SFD homogeneous system on any temporal scale. Similarly for the MIA in Fig. \ref{fig:HetR25Collapse}.

Concerning both Figs. \ref{fig:HomR25Collapse} and \ref{fig:HetR25Collapse}, the WFA FPTD also produces the required short-time concave structure and lies closer to the simulated data than the MIA (remembering the log-log scale), but still does not agree with it. In the long-time, as visible through Eq. (\ref{eq:SE H<1/3}) and Fig. \ref{fig:HomR25Collapse}, we see that although the power-law structure of $f_{\textrm{WF}}(t)$ does have a closer gradient than the $f_{\textrm{MI}}(t)$, we can still conclude that clearly the $f_{\textrm{WF}}(t)$  does not, again, agree with the simulated data on this scale.

\begin{figure}[!ht]
\centering
\includegraphics[scale=1.2]{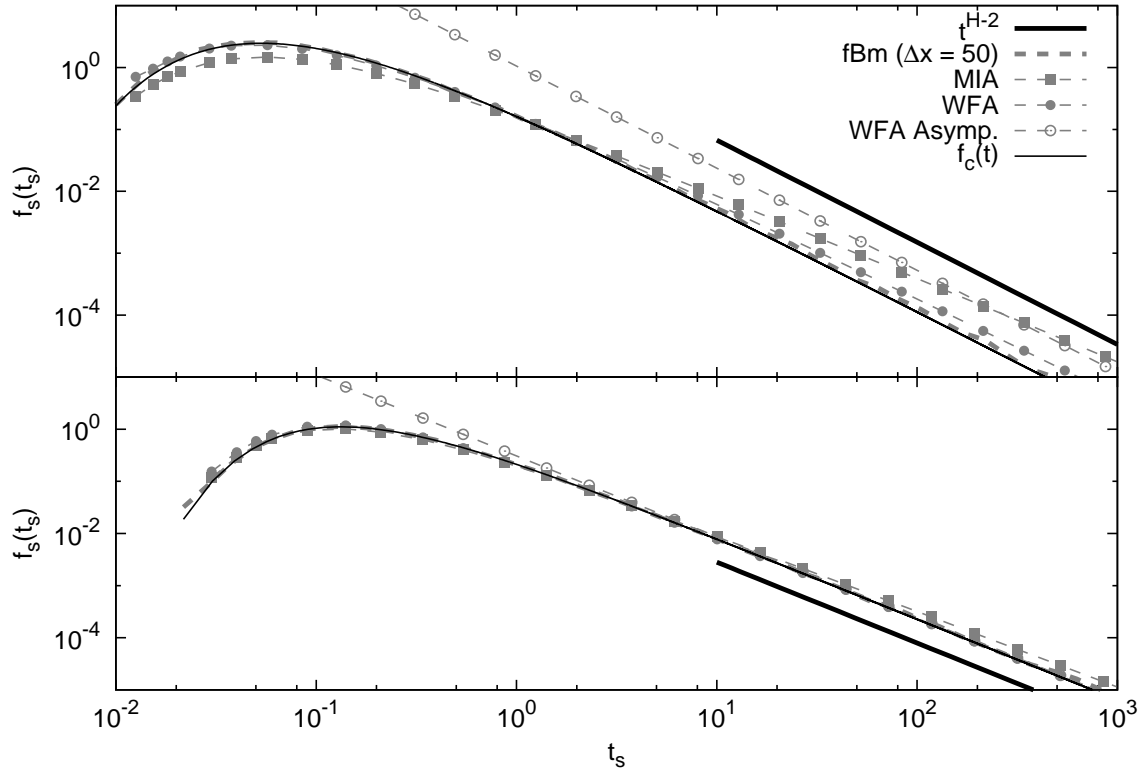} 
\caption{Collapsed plot of FPTD of sub-diffusive fBm with $H=0.35$ (top), $H=0.45$ (bottom) (simulation parameters: Table \ref{table:fBm_params}). Both panels show that the WFA has the correct heavy-tail gradient (compare to Molchan's prediction). The fBm also agrees with Eq. (\ref{eq:Molchan}), as the $f_{\textrm{c}}(t)$ tail is fixed $(H-2)$; and the data is modeled well by our conjecture, Eq. (\ref{eq:fptGuess}). Our conjecture models all time scales well, see Table \ref{table:fBm_params} for quantitative details.\label{fig:fBmH45Collapse}}
\end{figure}

{\bf fBm and WFA:} When one considers sub-diffusive fBm, under the criterion: $1/3\leq H<1/2$, we see, via Fig. \ref{fig:fBmH45Collapse} (where $H=0.35$, top, $H=0.45$, bottom), that the WFA agrees well with the fBm data: in the early time regime, and in the long time, has the correct gradient and a power-law pre-factor becomes exact as $H\rightarrow1/2$ (Brownian motion), as it should (and so to does the MIA in this limit).

\begin{figure}[!ht]
\centering
\includegraphics[scale=1.2]{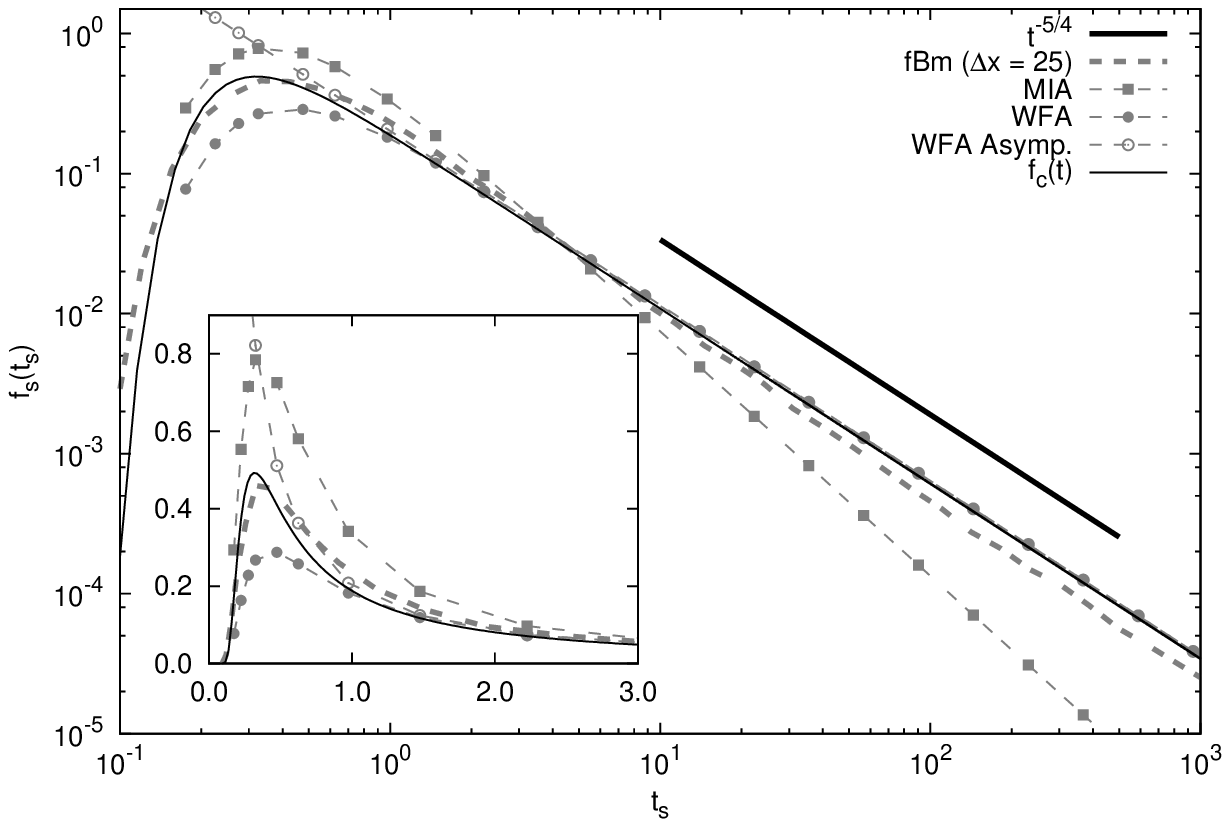} 
\caption{Collapsed plot of FPTD for super-diffusive fBm, $H=3/4$, see Table \ref{table:fBm_params}. Both the WFA and the fBm data only agree with Eq. (\ref{eq:Molchan}) in the long time, as expected. Using the conjecture, Eq. (\ref{eq:fptGuess}), keeping $H$ fixed, the 2 remaining degrees of freedom ($\gamma$, $\beta$, see Table \ref{table:fBm_params}) cannot account for the discrepancy seen between the our conjectured FPTD and the fBm data. It is apparent that the MIA fails on all time scales. INSET: Collapsed plot on linear axes illustrate short-time dynamics (fBm data as crosses).\label{fig:fBmH75Collapse}}
\end{figure}

For super-diffusive fBm, $H>1/2$, our simulated results, with $H=0.75$, except prediction of the correct long-time power-law exponent, show poor agreement with the WFA approximation, on every scale - see Fig. \ref{fig:fBmH75Collapse}. For simulation and fitting data see Table \ref{table:fBm_params}.

{\bf Conjecture results: } The viability of the trial function, Eq. (\ref{eq:fptGuess}), was measured with comparison to simulated data (both SFD and fBm; Tables \ref{table:GuessFitHom} and \ref{table:fBm_params}, respectively). 

In the homogeneous system,  Fig. \ref{fig:HomR25Collapse}, the power-law exponent, is fixed to the theoretical result: $(H-2)$ ; where $H$ is taken from the literature in the form of Eq. (\ref{eq:MSDboxHom}), leaving only two parameters, $\gamma$ and $\beta$; to be fitted (see Appendix \ref{sec:DataRed}), resulting in Table \ref{table:GuessFitHom} \cite{fs_vs_ts}.
\begin{table}[ht]
\caption{Homogeneous $f_{\textrm{c}}(t)$ fit parameters. $\hat{\chi}^2$ is the normalized Chi-squared parameter. Raw data placed into 30 natural log-bins (see Appendix \ref{sec:DataRed}).}
 \centering
\begin{tabular}{|c|c|c|c|}
	\hline
$x_c$ & $\gamma $ & $\beta$ & $\hat{\chi}^2$\\
	\hline
25 &$0.72\pm0.03$& $0.96\pm 0.02$& 1.49 \\
40&$0.67\pm0.03$& $0.97\pm0.02$& 1.64\\
50& $0.64\pm0.03$& $0.98\pm0.02$&1.61\\
75& $0.60\pm0.02$& $1.00\pm0.02$&1.33\\
	\hline
\end{tabular}
\label{table:GuessFitHom}
\end{table}
The low $\hat{\chi}^2$ value, shown in Table \ref{table:GuessFitHom}, indicates that our proposed form for the FPTD is indeed an accurate one. Further, the results in the $\beta$ column of Table \ref{table:GuessFitHom}, leads us to propose that $\beta = 1$ (as is for the Brownian case), \emph{i.e.} the FPTD is well approximated by the propagator, Eq. (\ref{eq:GeneralPDF}), multiplied by a power-law pre-factor chosen to reproduce Molchan's long-time relation.

For the heterogeneous case, the simulation data have been subjected to the same treatment as applied to the homogeneous system data; yielding poor results. The full functional form of the heterogeneous system is more complicated and not amenable to explanation by our simple conjecture. This complexity, as stated previously, arises predominantly through creating an entirely new heterogeneous population for each simulation (heterogeneity-averaged). Each different population, three of which are illustrated in the inset of Fig. \ref{fig:HetR25Collapse}, has a different short-time FPTD structure relative to other populations as well as a different pre-factor in the power-law tail. Thus, in short, the FPTD for heterogeneous SFD shows the same kind of lack of self-averaging as displayed in the MSD for this type of system \cite{lomholt-2010}. In particular, the pre-factor in the long-time asymptotics is a random quantity dependent on the particular realization of the friction constant population. 

Upon inspection of Figs. \ref{fig:HomR25Collapse} to \ref{fig:fBmH45Collapse}, it is easily seen that the conjecture, Eq. (\ref{eq:fptGuess}), works well at modeling the FPTD dynamics of fBm for sub-diffusive and diffusive systems. When considering super-diffusive motion, Fig. \ref{fig:fBmH75Collapse}, it is clear that the conjecture is not adequate to deal with these dynamics. Quantitatively, these statements are displayed in Table \ref{table:fBm_params}.
\begin{table}[ht]
\caption{fBm Simulation and Fit Parameters. Each simulation has: $C=5$; $t_{\textrm{stop}}=10^7$; $N=6\times10^4$. Raw data placed in 50 natural log-bins, fitted to parameters $(\gamma,\beta)$ (see Appendix \ref{sec:DataRed}) to test the applicability of our simple conjecture, Eq. (\ref{eq:fptGuess}). $\hat{\chi}^2$ is the normalized Chi-squared parameter.}
 \centering
\begin{tabular}{|c|c|c|c|c|}
	\hline
$H$ & $\Delta x$ & $\gamma$& $\beta$ & $\hat{\chi}^2$\\
	\hline
1/6  & 25 & $0.726 \pm 0.004$ & $1.00\pm 0.02$ & $1.12$ \\
1/6  & 75 & $0.556 \pm 0.004$ & $1.004 \pm 0.002$ & $1.21$ \\
1/4  & 50 & $0.639 \pm 0.005$ & $0.987 \pm 0.003$ & $1.70$\\
1/4  & 100 & $0.595 \pm 0.005$ & $0.984 \pm 0.003$ & $1.65$ \\
7/20  & 50 & $0.667 \pm 0.006$ & $0.958 \pm 0.005$ & $3.16$ \\
9/20  & 50 & $0.595 \pm 0.007$ & $0.973 \pm 0.006$ & $0.985$  \\
3/4  & 200 & $0.0862 \pm 0.0004$ & $1.70 \pm 0.03$& $55.7$ \\
	\hline
\end{tabular}
\label{table:fBm_params}
\end{table}
\section{Summary and Discussion\label{sec: conclu}}
We investigated the first passage time densities (FPTD) of a tracer particle in a single-file system with two different population types; homogeneous (all having the same diffusion constant), and heterogeneous (friction constants drawn from a heavy-tail power-law distribution). Along side, the fractional Brownian motion (fBm) FPTD was investigated.

Theoretically, two methods were used to approximate the full functional form for FPTD analytically: the Method of Images approximation, Eq. (\ref{eq:MIApprox}), and the Willemski-Fixman approximation, Eq. (\ref{eq: series-exp}). Moreover, a conjectured form for the FPTD, Eq. (\ref{eq:fptGuess}), was introduced. Numerically, the SFD and fBm systems were simulated stochastically with the Gillespie-type algorithm presented in Ref. \cite{ambjornsson-2008-129}, and spectral approach \cite{Dieker2003}, respectively.

Our main conclusions are:
\begin{itemize}
 \item The MIA derived here, and previously used in the literature, does not approximate the FPTD at any temporal scale.
 \item With the use of Mellin transforms we have found an exact result for the full FPTD within the Willemski-Fixman approximation. To convert the inverse FPTD from Mellin frequency space two methods were used: the series expansion approach and the Weyl fractional derivative approach. The WFA does not agree with simulated the SFD FPTD (homogeneous and heterogeneous) nor the fBm FPTD for $H<1/3$. For sub-diffusive and Brownian motion, \emph{i.e.} $1/3\leq H<1/2$, the WFA approximates the FPTD well at all times, including the theoretical long-time slope, Eq. (\ref{eq:Molchan}), (but does not predict the correct pre-factor) becoming exact when $H=1/2$. In the super-diffusive regime, $H\in(1/2,1)$, the WFA manages to capture only the long-time power-law exponent correctly.
 \item We show through simulations that the FPTD for the homogeneous SFD system is equivalent to the fBm FPTD with $H=1/4$, for all times, in the correct scaling.
 \item We find that the FPTD for a heavy-tailed heterogeneous SFD system and the corresponding fBm share the same power-law exponent for long times (given the correct scaling, Eq. (\ref{eq:MSDboxHet})). The general heterogeneous SFD FPTDs show a lack of self-averaging. 
 \item A simple conjecture, Eq. (\ref{eq:fptGuess}), is proposed and adequately show to model the full functional form of the fBm FPTD for $H\leq 1/2$; and also that of the homogeneous SFD FPTD. Eq. (\ref{eq:fptGuess}) only captures the asymptotic power-law exponent for heterogeneous SFD systems and super-diffusive systems where $H>1/2$.

\end{itemize}

Our work presented herein also pertains to research on FPTs in many-body ``swarm'' systems; such as the recent work by Mej\'{\i}a-Monasterio \emph{et al.} \cite{Mejia-Monasterio2011}. They have investigated the FPT for a search by a swarm of independent searchers. Our results initialize the generalization of this investigation, and further paves the way for the understanding of FPTs in general interaction, non-independent, many-body swarm systems.

Whereas there is a wealth of knowledge about first passage problems for Markovian systems \cite{Redner2001}, our understanding of the corresponding problem for non-Markovian dynamics is far from complete. This study shows quantitatively the limitations of two commonly used approximations, provides extensive simulation results for homogeneous and heterogeneous system, and a conjectured form for the FPTD for the fBm, thereby providing headway into future studies.

\begin{acknowledgments}
 We would like to acknowledge the following persons for their insightful discussions on various aspects of this investigation: Michael Lomholt, Ludvig Lizana, Michaela Schad, and Sigur\dh ur \AE. J\'onsson. Computer time was provided by LUNARC at Lund University.
\end{acknowledgments}

\appendix
\section{WFA Derivation\label{sec:WFA}}
The canonical method for solving the Willemski-Fixman problem, Eq. (\ref{eq:renewal}), is to use the Laplace transform (operator $L$) to take advantage of the convolution property of this transform. However, this transform for the function: $L\left[\exp\left(-Vt^{-r}\right)\right]$ with respect to $t$; does not exist in closed form for arbitrary $r$. In here, we instead therefore use the Mellin transform technique.

Beginning from the integral Eq. (\ref{eq:renewal}), plugging in the propagator, Eq. (\ref{eq:GeneralPDF}); multiplying both sides by $t^H$, we write
\begin{equation}\label{eq:formalMellin}
 \exp\left(-\sigma t^{-2H}\right)=\int_0^\infty F(x_c,t'|x_0)G(t/t')\frac{dt'}{t'},
\end{equation}
with $F(x_c,t'|x_0)\equiv t'f(x_c,t'|x_0)$, where $\sigma=(\Delta x^2)/(4C)$ as before, and
\[
 G(t/t')\equiv \left(\frac{t}{t'}\right)^H\left(\frac{t}{t'}-1\right)^{-H}\Theta\left(\frac{t}{t'}-1\right).
\]
where $\Theta(t)$ is the Heaviside function. We now can solve Eq. (\ref{eq:formalMellin}), using the Mellin Transformation (see \cite{IntTransBook} for more details). This transform is defined as
\begin{equation}
 \hat{g}(p)= M\left[g(t)\right]\equiv\int_0^{\infty}t^{p-1}g(t)dt.
\end{equation}
Using the following identities:
\[
 M\left[\int_0^\infty y(t')g\left(\frac{t}{t'}\right)\frac{dt'}{t'}\right]=\hat{y}(p)\hat{g}(p),
\]
\begin{equation}\label{eq:MellinConvol}
 M\left[g\left(t^{-w}\right)\right]_{w>0}=\frac{1}{w}\hat{g}\left(\frac{-p}{w}\right),
\end{equation}
\[ M\left[e^{-at}\right]_{a>0}=a^{-p}\Gamma(p).
\]
the formal solution to Eq. (\ref{eq:formalMellin}) becomes 
\begin{equation}\label{eq:ConvolFormalSol}
 \frac{1}{2H}\sigma^{p/2H}\Gamma\left(\frac{-p}{2H}\right)=\hat{F}(x_c,p|x_0)\hat{G}(p),
\end{equation}
where
\[
 \hat{G}(p)=\int_0^{\infty}t^{p-1}t^{H}(t-1)^{-H}\Theta(t-1)dt=\int_1^{\infty}t^{p+H-1}(t-1)^{-H}dt=B(-p,1-H).
\]
$B(z,w)$ is the Beta function \cite{ABST}, where $B(z,w)=\left[\Gamma(w)\Gamma(z)\right]/\Gamma(z+w)$. Following this route, one finds that Eq. (\ref{eq:ConvolFormalSol}) becomes
\begin{equation}
 \hat{F}(x_c,p|x_0)=\frac{\sigma^{p/2H}}{2H\Gamma(1-H)}\frac{\Gamma(1-H-p)\Gamma\left(\frac{-p}{2H}\right)}{\Gamma(-p)},
\end{equation}
for $\textrm{Re}(p)<0$ \cite{FunStrip}. Using the fact that $M\left[F(t)t^{-1}\right]=\hat{F}(p-1)$,
(remembering $F(x_c,t|x_0)=tf(x_c,t|x_0)$) one finally gets
\begin{equation}\label{eq:MellTransFormal}
 \hat{f}(x_c,p|x_0)=\frac{\sigma^{(p-1)/2H}}{2H\Gamma(1-H)}\frac{\Gamma(2-H-p)\Gamma\left(\frac{1-p}{2H}\right)}{\Gamma(1-p)}
\end{equation}
for $\textrm{Re}(p)<1$ , which gives the closed form expression for the Mellin transform of the first passage time density $f(x_c,t|x_0)$, within the WFA. 

\subsection{Series Expansion Approach}

From Eq. (\ref{eq:MellTransFormal}), and the formal Mellin-inversion formula, we have:
\begin{equation}\label{eq:invertFPT}
 f(x_c,p|x_0)=\frac{1}{2H\Gamma\left(1-H\right)}\frac{1}{2\pi i}\int_{c-i\infty}^{c+i\infty}t^{-p}\sigma^{\frac{p-1}{2H}}\Gamma\left(2-H-p\right)\frac{\Gamma\left(\frac{1-p}{2H}\right)}{\Gamma\left(1-p\right)}dp.
\end{equation}
We can deform the original contour where $c$ is chosen such that the integration is over the fundamental strip (here $\textrm{Re}(p)<1$), so that it becomes a square, side length $R$, which encompasses the fundamental strip, encloses all poles in the domain $\textrm{Re}(p)>1$ and has a negative orientation (anti-clockwise) - see \cite{Hughes}, Appendix A therein. We can then use the residue theorem to compute Eq. (\ref{eq:invertFPT}):
\begin{equation}\label{eq:Sum_res}
 f(x_c,p|x_0)=\frac{-1}{2H\Gamma\left(1-H\right)}\sum_{\textrm{Residues}} \textrm{Res}\left[t^{-p}\sigma^{\frac{p-1}{2H}}\Gamma\left(2-H-p\right)\frac{\Gamma\left(\frac{1-p}{2H}\right)}{\Gamma\left(1-p\right)}\right],
\end{equation}
where we used the fact that the integral along the upper, lower, and right parts of the square vanish as $R\rightarrow\infty$. Note the minus sign in the pre-factor due to anti-clockwise contour. $\Gamma(p)$ has simple poles at $p=0,-1,-2,...$; and the function inside the square brackets on the RHS of Eq. (\ref{eq:Sum_res}) therefore has (potential) poles at 
\begin{equation}\label{eq:p_n}
  p_n = 2-H+n;\quad n=0,1,2,...
\end{equation}
\begin{equation}\label{eq:p_m}
 \tilde{p}_m = 1+2Hm;\quad m=0,1,2,...
\end{equation}
To proceed we need to know the behavior of $\Gamma\left[\frac{\alpha-z}{\beta}\right]$ close to $z_m = \alpha-m\beta$. We have that for a simple pole the residue is calculated as:
\[
 \textrm{Res}\left[\Gamma \left(\frac{\alpha-z}{\beta}\right)\right]_{z=z_m}=\lim_{z\to z_m}(z-z_m)\Gamma \left(\frac{\alpha-z}{\beta}\right),
\]
after some manipulation, by inserting $z_m=\alpha+m\beta$ and using $\Gamma(z) = z^{-1}\Gamma(z+1)$, we find that
\begin{equation}\label{eq:resGammaFrac}
 \textrm{Res}\left[\Gamma \left(\frac{\alpha-z}{\beta}\right)\right]_{z=z_m}=\frac{-\beta}{m\beta}\frac{-\beta}{(m-1)\beta}\frac{-\beta}{(m-2)\beta}...(-\beta)\Gamma(1)=\frac{(-1)^{m+1}}{m!}\beta.
\end{equation}
Noting that $1/\Gamma(z)$ has no poles, and combining Eqs. (\ref{eq:Sum_res}) and (\ref{eq:resGammaFrac}) we get
\begin{align}
f(x_c,t|x_0)&&&=&\frac{-1}{2H\Gamma\left(1-H\right)}\bigg{[}\sum_{n=0}^{\infty}t^{-p_n}\sigma^{\frac{p_n-1}{2H}}\frac{(-1)^{n+1}}{n!}\frac{\Gamma\left(\frac{1-p_n}{2H}\right)}{\Gamma\left(1-p_n\right)} \nonumber\\
&&& &+\sum_{m=1}^{\infty}t^{-\tilde{p}_m}\sigma^{\frac{\tilde{p}_m-1}{2H}}\frac{(-1)^{m+1}}{m!}2H\frac{\Gamma\left(2-H-\tilde{p}_m\right)}{\Gamma\left(1-\tilde{p}_m\right)}\bigg{]}\nonumber
\end{align}
substituting in $K(t)$, see Eq. (\ref{eq:WFK}), and the values for $p_n$ and $\tilde{p}_m$, one arrives at the exact expression for the first passage time within the WFA, which can be used to numerically evaluate the FPTD:

\begin{align}\label{eq:GenerealSeriesExpan}
f(x_c,t|x_0)&&&=&\frac{\sigma^{-1/2H}}{\Gamma\left(1-H\right)}\bigg{[}\frac{1}{2H}\left[K(t)\right]^{\left(2-H\right)}\sum_{n=0}^{\infty}\left[K(t)\right]^{n}\frac{(-1)^{n}}{n!}\frac{\Gamma\left(\frac{H-1-n}{2H}\right)}{\Gamma\left(H-1-n\right)} \nonumber \\
&&&&+K(t)\sum_{m=1}^{\infty}\left[K(t)\right]^{2Hm}\frac{(-1)^{m}}{m!}\frac{\Gamma\left(1-H-2Hm\right)}{\Gamma\left(-2Hm\right)}\bigg{]}
\end{align}
We point out that we above assumed simple poles, \emph{i.e.} that $p_n\neq \tilde{p}_m$. Using Eqs. (\ref{eq:p_n}) and (\ref{eq:p_m}), we see that we therefore require $H\neq(n+1)/(2m+1)$ for Eq. (\ref{eq:GenerealSeriesExpan}) to be valid. 

Now, an immediate check for the calculation of Eq. (\ref{eq:GenerealSeriesExpan}) is performed when $H=1/2$, \emph{i.e.} the system's dynamics is Brownian motion. Then
\[
f(x_c,t|x_0)=\frac{\sigma^{-1}}{\Gamma\left(\frac{1}{2}\right)}\left(\left[K(t)\right]^{3/2}\sum^{\infty}_{n=0}\left[K(t)\right]^{n}\frac{(-1)^n}{n!} +K(t)\sum^{\infty}_{m=1}\left[K(t)\right]^{m}\frac{(-1)^m}{m!}\frac{\Gamma\left(\frac{1}{2}-m\right)}{\Gamma\left(-m\right)}\right),
\]
where we note the sum of $n$ term is the power series of an exponential function, and that, in the second sum, over $m$,  $1/\Gamma(-m)=0$ for an integer $m\geq0$. Substituting in the functional value for $K(t)$, we get
\[
 f(x_c,t|x_0)=\frac{\sigma^{1/2}}{\sqrt{\pi}}t^{-3/2}\exp\left(\frac{-\sigma}{t}\right),
\]
as required \cite{Redner2001}.

\subsection{Asymptotic Time Limit: $K(t)\rightarrow0$}
In the long-time limit $n=0$ term and the $m=1$ term in Eq. (\ref{eq:GenerealSeriesExpan}) give the dominant contribution to the FPTD, namely
\begin{equation}\label{eq:series_limit}
f(x_c,t|x_0)\sim\frac{\sigma^{-1/2H}}{\Gamma\left(1-H\right)}\left(\left[K(t)\right]^{2-H}\frac{1}{2H}\frac{\Gamma\left(\frac{H-1}{2H}\right)}{\Gamma\left(H-1\right)} -\left[K(t)\right]^{1+2H}\frac{\Gamma\left(1-3H\right)}{\Gamma\left(-2H\right)}\right).
\end{equation}
We see that the exponents in Eq. (\ref{eq:series_limit}) are equal if $H=1/3$. Thus this obliges us to make three distinct cases for the FPTD: when $H>1/3$ we use the reflection formula for $\Gamma-$functions and arrive at Eq. (\ref{eq:SE H>1/3}); in the case that $H<1/3$, we find Eq. (\ref{eq:SE H<1/3}), using $\Gamma(1+z)=z\Gamma(z)$. For the case that $H=1/3$ (or more generally of the form $H^{*}_{nm}=(n+1)/(2m+1)$), a double pole exists and certain care is needed.

{\bf Double Pole:} When $H^{*}_{nm}=(n+1)/(2m+1)$, \emph{e.g.} $H=1/3$, one or several double poles exist and should be dealt with accordingly. To see so, if we substitute  $H=1/3$ into both Eq. (\ref{eq:SE H>1/3}) and Eq. (\ref{eq:SE H<1/3}) we find a diverging pre-factor which is proportional to $1/\left(H-1/3\right)$, which is not correct as $f(x_c,t|x_0)$ is bounded, namely: $\int^{\infty}_{-\infty}f(x_c,t|x_0)dt=1$. In order to resolve the double pole issue we need to revisit Eq. (\ref{eq:Sum_res}).

The residue for a pole of order 2 is computed as 
\begin{equation}\label{eq:N2_res}
\textrm{Res}\left[F(z)\right]=\lim_{z\to z_m}\frac{d}{dz}\left[(z-z_m)^2F(z)\right].
\end{equation}
Considering Eq. (\ref{eq:Sum_res}), we note that we need to evaluate - following the same steps that led to Eq. (\ref{eq:Sum_res}) - (remembering: $z_n = \eta +n$, $n=0,1,2,...$ and $\tilde{z}_m=\alpha+m\beta$),
\begin{align}\label{eq:double_pole_gamma}
\textrm{Res}\left[\Gamma(\eta-z)\Gamma\left(\frac{\alpha-z}{\beta}\right)\right]_{z=\tilde{z}_m=z_n}&&&=&(-\beta)^{m+1}(-1)^{n+1}\frac{d}{dz}\bigg{[}\nonumber\\
&&&&\frac{\Gamma\left(\frac{\alpha-\tilde{z}+(m+1)\beta}{\beta}\right)}{(\tilde{z}-\alpha)(\tilde{z}-\alpha-\beta)\cdots(\tilde{z}-\alpha-(m-1)\beta)}\nonumber\\
&&&\times&\frac{\Gamma\left(\eta-z+(n+1)\right)}{(z-\eta)(z-\eta-1)\cdots(z-\eta-(n-1))}\bigg{]}
\end{align}
Eqs. (\ref{eq:Sum_res}) and (\ref{eq:double_pole_gamma}) allow us to compute $f(x_c,t|x_0)$ (in principle), also for $n$ and $m$ values satisfying  $H^{*}_{nm}=(n+1)/(2m+1)$. For a given $(n,m)$ from Eq. (\ref{eq:double_pole_gamma}), we see that the residue for the double pole involves $2+n+m$ terms. Let us now limit our derivation to the analysis of $H=1/3$ and long times (where $n=0$ and $m=1$), then Eq. (\ref{eq:double_pole_gamma}) becomes
\begin{align}
\textrm{Res}\left[\Gamma(\eta-z)\Gamma\left(\frac{\alpha-z}{\beta}\right)\right]&&&=&\beta^2\frac{\Gamma\left(\frac{\alpha-\tilde{z}}{\beta}+2\right)\Gamma\left(\eta-z+1\right)}{\tilde{z}-\alpha}\nonumber\\
&&&\times&\left[\frac{1}{\tilde{z}-\alpha}+\frac{1}{\beta}\frac{\Gamma'\left(\frac{\alpha-\tilde{z}}{\beta}+2\right)}{\Gamma\left(\frac{\alpha-\tilde{z}}{\beta}+2\right)}+\frac{\Gamma'\left(\eta-z+1\right)}{\Gamma\left(\eta-z+1\right)}\right].\nonumber
\end{align}
For the investigation in this report, we have $\beta =2H$; $\tilde{z}_m=\tilde{p}_m=1+2Hm\rightarrow1+2H$; and $\alpha = 1$; therefore $z_n=p_n=2-H$, and finally $\eta=2-H$. Substituting these values in above, and setting $H=1/3$ we find
\[
  \textrm{Res}\left[\Gamma(2-H-z)\Gamma\left(\frac{1-z}{2H}\right)\right]=1+\frac{5}{3}\gamma_{E}
\]
where $\left(\Gamma'(x)\right)/\left(\Gamma(x)\right)|_{x=1}=\gamma_E$ (Euler's constant) \cite{ABST}. Eq. (\ref{eq:Sum_res}) now becomes (with $p=p_0=2-H$) Eq. (\ref{eq:SE H=1/3}), as given in the main text.

\subsection{Weyl Fractional Derivative Approach}

Let us now derive Eq. (\ref{eq:WFf}) in the main text. To obtain the FPTD, one has to invert the transform in Eq. (\ref{eq:MellTransFormal}). The inverse Mellin transform is defined as \cite{IntTransBook} 
\[
 f(t)\equiv\frac{1}{2\pi i}\int_{c-i\infty}^{c+i\infty}\hat{f}(p)t^{-p}dp,
\]
where $a<c<b$ and $\langle a,b\rangle$ is the the fundamental strip; in this case, $\langle -\infty,1\rangle$. Transforming the inverse integral of Eq. (\ref{eq:MellTransFormal}) to the variable $\tilde{p}=2-H-p$ we obtain:
\begin{equation}\label{eq:CloseF}
 f(x_c,t|x_0)=\frac{\sigma^{-1/2H}}{\Gamma(1-H)}\left[K(t)\right]^{2-H}\phi\left[K(t)\right],
\end{equation}
\begin{equation}\label{eq:funkyPhi}
 \phi(z)=\frac{1}{2H}M^{-1}\left[\frac{\Gamma(p)\Gamma\left(\frac{p-(1-H)}{2H}\right)}{\Gamma\left[p-\left(1-H\right)\right]},z\right],
\end{equation}
\begin{equation}
 K(t)=\frac{\sigma^{1/2H}}{t}.
\end{equation}
The quantity $\phi(z)$ can be written in terms of H-functions or Fox-functions, but from \cite{IntTransBook}, Eq. (8.5.23) one has
\begin{equation}
 M^{-1}\left[\frac{\Gamma(p)}{\Gamma(p-\beta)}\hat{g}(p-\beta),z\right]=W^{\beta}\left[g(z)\right]\equiv\frac{(-1)^n}{\Gamma(n-\beta)}\frac{d^n}{dz^n}\int_z^\infty(z'-z)^{n-\beta-1}g(z')dz',
\end{equation}
where $W^{\beta}$ is the Weyl fractional derivative and $n$ is the smallest integer such that $n-\beta>0$ \cite{IntTransBook}. In this case - Eq. (\ref{eq:funkyPhi}) - one has $\beta=1-H$ which implies $0<\beta<1$, so $n=1$ and $\hat{g}(p)=\Gamma\left(p/2H\right)$. Therefore $g(z)=M^{-1}\left[\Gamma\left(p/2H\right),z\right]=2H\exp\left(-z^{2H}\right)$, and 
\begin{align}
 \phi(z)&=W^{1-H}\left[\exp\left(-z^{2H}\right)\right]\\ \nonumber
&=\frac{-1}{\Gamma(H)}\frac{d}{dz}\int_z^{\infty}(z'-z)^{-(1-H)}\exp\left[-(z')^{2H}\right]dz'\\ \nonumber
&=\frac{2H}{\Gamma(H)}\int_0^{\infty}y^{-(1-H)}(z+y)^{2H-1}\exp\left[-(z+y)^{2H}\right]dy\nonumber
\end{align}
where, in the last step, we changed variables $y=z'-z$.
The diverging factor $y^{-(1-H)}$ in the integrand is inconvenient for numerical evaluation of the integral, so to remedy this, we integrate by parts to get
\begin{equation}\label{eq:endPhi}
 \phi(z)=\frac{2}{\Gamma(H)}\int_0^{\infty}y^{H}\left[2H(z+y)^{2H}-(2H-1)\right](z+y)^{2H-2}\exp\left[-(z+y)^{2H}\right]dy.
\end{equation}
Using the reflection formula for Gamma functions \cite{ABST}, $\Gamma(z)\Gamma(1-z)=\pi/\sin(\pi z)$, (note: $0<z<1$) we finally get - by inserting Eq. (\ref{eq:endPhi}) into Eq. (\ref{eq:CloseF}),  Eq. (\ref{eq:WFf}) given in the main text. Careful considerations show that the integration by parts is only valid for $K(t)\neq0$ when $H\leq1/3$ providing limits for the validity of Eq. (\ref{eq:WFf}). 


\section{MIA Derivation\label{sec:MIA}}
Given the result of the PDF from $\S$ \ref{sec: Approx} - Eq. (\ref{eq:GeneralPDF}), we have the PDF in the presence of an absorption point located at $x_c$,
\[
 P_{\textrm{MI}}(x,t|x_0)=P(x,t|x_0) - P(x,t|2x_c-x_0).
\]
This combination of PDFs enforce the absorbing boundary condition $P_{\textrm{MI}}(x=x_c,t|x_0)=0$. Using this result, the Survival probability is $ S_{\textrm{MI}} (t)= S_{\textrm{A}}(t)-S_{\textrm{B}}(t)$, where
\begin{equation}
 S_{\textrm{A}}(t) =\int_{-\infty}^{x_c}P(x,t|x_0)dx;\quad S_{\textrm{B}}(t) =\int_{-\infty}^{x_c}P(x,t|2x_c-x_0)dx.
\end{equation}
Making use of the Error function, $\mbox{Erf}[z]$ \cite{ABST}, one finds that 
\begin{equation}
 S_{\textrm{A}} =\frac{1}{2}\left[1+\mbox{Erf}\left(\frac{x_c-x_0}{\sqrt{4Ct^{2H}}}\right)\right]; \quad S_{\textrm{B}} =\frac{1}{2}\left[1+\mbox{Erf}\left(\frac{x_0-x_c}{\sqrt{4Ct^{2H}}}\right)\right].
\end{equation}
The relationship between the FPTD and the Survival probability is $f_{\textrm{MI}}(t) = -\frac{\partial}{\partial t}S_{\textrm{MI}}(t)$ \cite{Redner2001}, this leads to the final result predicted by the MIA for a non-Markovian system - Eq. (\ref{eq:MIApprox}) - in the main text.

\section{Data Reduction and Fitting\label{sec:DataRed}}
The simulated data presented here was subjected to the following data reduction pipeline: first passage times are culminated from simulations then the data are log-binned and plotted as histograms. When fitting of histograms is required, the boot-strap method \cite{Efron-77} is used to estimate the error of each bin. In this method one creates a large number (in our case $10^3$) ``synthetic'' raw data sets from the original raw data (the vector/set containing the FPTs). Each synthetic set is accomplished by selecting, at random (from a uniform distribution), an entry from the original set, \emph{with replacement}, until the synthetic set being created has the same number of entries as the original. This is repeated until the desired amount of synthetic data sets is achieved. Each synthetic set is then binned into bins of the same size as the original data set, and the variance for each bin (acquired from binning a large number of synthetic data sets) is then used as the error for the original bin. This error is subsequently used to weight the least-squares fitting.

After boot-strapping, the Levenberg-Marquardt regression (LMA) is employed to find the best fit \cite{ChiMin, ChanceEncounters}. When fitting the FPTD conjecture \cite{fs_vs_ts}, the LMA was employed to minimize the \emph{Chi-squared} value, $\chi^2$. To be more explicit, the normalized Chi-squared value, $\hat{\chi}^2$, was minimized as it is more easily interpreted (namely without the use of tables). This is just: $\chi^2/\textrm{dof}$, where $\textrm{dof}$ is the degrees of freedom of a fit. If this value lies between $0.5<\hat{\chi}^2<1.5$ the model is assumed to be an accurate explanation for the functional form of the data.



\begin{thebibliography}{74}
\bibitem [{Flu()}]{Fluc_diss_note}%
  \BibitemOpen
  \href@noop {} {}\bibinfo {note} {In fact, the relation between the noise
  correlation function and the memory kernel for fractional Langevin dynamics
  is a direct consequence of the fluctuation-dissipation theorem.}\BibitemShut
  {Stop}%
\bibitem [{App()}]{Approx_sim}%
  \BibitemOpen
  \href@noop {} {}\bibinfo {note} {Note on WF and MI Approximations: When
  numerically calculating the approximations seen in Figs.
  \ref{fig:HomR25Collapse}-\ref{fig:fBmH75Collapse}, we set $C$,
  $t_{\rm{stop}}$, and $\Delta x$ to those values of the fBm simulation;
  see Table \ref{table:fBm_params}.}\BibitemShut {Stop}%
\bibitem [{fBm(){\natexlab{a}}}]{fBm_hom_conjec_fit}%
  \BibitemOpen
  \href@noop {} {}\bibinfo {note} {The fBm data for both
  $\Delta x$ show in Fig. \ref{fig:HomR25Collapse}, were also fitted with the
  conjecture, and whose results are quantified in Table \ref{table:fBm_params};
  but the curves are not plotted, as this would obfuscate the more important
  aspects of the said figure.}\BibitemShut {Stop}%
\bibitem [{fBm(){\natexlab{b}}}]{fBm_overlap_hom}%
  \BibitemOpen
  \href@noop {} {}\bibinfo {note} {We require
  $|x_c|\gg|x_0|$ for the scaling we have used to be valid. Therefore we have
  overlapped two fBm FPTD data sets in Fig. \ref{fig:HomR25Collapse}, where
  FPTD with $\Delta x =100$ is plotted for $t<0.2$, and FPTD with $\Delta x=50$
  is shown for $0.1<t$. The larger $\Delta x$ in the short-time range is a more
  realistic representation of those dynamics in this scaling, but to show that
  the scaling works in the long time, a smaller $\Delta x$ is used to capture
  the dynamics completely. The overlap suggests the scaling collapses both data
  on top of each other.}\BibitemShut {Stop}%
\bibitem [{fBm(){\natexlab{c}}}]{fBm_overlap_het}%
  \BibitemOpen
  \href@noop {} {}\bibinfo {note} {For the same reasons
  stated in \cite{fBm_overlap_hom} we plot the data from two $\Delta
  x\in\{25,75\}$ for fBm, in Fig. \ref{fig:HetR25Collapse}, where the overlap
  time range is $1.5\times10^{-3}<t<2.5\times10^{-3}$. The conjecture fitted,
  and shown in this plot is the fitted function, Eq. (\ref{eq:fptGuess}), from
  each $\Delta x$ spliced together over the time frames that each fBm data set
  is plotted for.}\BibitemShut {Stop}%
\bibitem [{fs_()}]{fs_vs_ts}%
  \BibitemOpen
  \href@noop {} {}\bibinfo {note} {We actually fit the \emph{scaled} data to
  the \emph{scaled} conjecture:
  $f_{s}(t_s)=(2H\beta\gamma^{G})/(\Gamma\left(G\right)2^{G})t_s^{H-2}
  \exp\left(-\gamma2^{-\beta}t_s^{-2H\beta}\right)$, where $G=(1-H)/(2H\beta)$.
  One then notes that $f_s(t_s)$ is independent of $\Delta x$ and $C$, as
  required.}\BibitemShut {Stop}%
\bibitem [{Fun()}]{FunStrip}%
  \BibitemOpen
  \href@noop {} {}\bibinfo {note} {The Fundamental strip is the range of
  $p-$values for which the transform
  $M\left[e^{-at^{-\beta}}\right]=\int_0^{\infty}t^{p-1}e^{-at^{-\beta}}dt$
  exists. The exponential in the previous equation has the limits: 1 as
  $t\rightarrow\infty$ and 0 as $t\rightarrow 0$. The fundamental strip is
  therefore: $\textrm{Re}(p)<0$ or $\langle -\infty,0\rangle$.}\BibitemShut
  {Stop}%
\end{thebibliography}

%

\end{document}